\documentclass[aps,prl,twocolumn,superscriptaddress,showpacs, amsmath,amssymb]{revtex4-2}

\usepackage{graphicx}
\usepackage{dcolumn}
\usepackage{bm}
\usepackage{xcolor}

\usepackage{hyperref}

\begin{document}

\title{Finite-Frequency Topological Maxwell Modes in Mechanical Self-Dual Kagome Lattices}


\author{Hrishikesh Danawe}
\affiliation{Department of Mechanical Engineering, University of Michigan, Ann-Arbor, MI 48109-2125, United States}
\author{Heqiu Li}
\affiliation{Department of Physics, University of Michigan, Ann-Arbor, MI 48109-2125, United States}
\affiliation{Department of Physics, University of Toronto, Toronto, Ontario M5S 1A7, Canada}
\author{Kai Sun}
\affiliation{Department of Physics, University of Michigan, Ann-Arbor, MI 48109-2125, United States}
\author{Serife Tol}
\affiliation{Department of Mechanical Engineering, University of Michigan, Ann-Arbor, MI 48109-2125, United States}

\date{\today}

\begin{abstract}
In this Letter, an elastic twisted kagome lattice at a critical twist angle, called self-dual kagome lattice, is shown to exhibit peculiar finite-frequency topological modes which emerge when certain conditions are satisfied. These states are topologically reminiscent to the zero energy (floppy) modes of Maxwell lattices but they occur at a finite frequency in the band gap of self-dual kagome lattice. Thus, we present a completely new class of topological modes which share similarities with both the zero frequency floppy modes in Maxwell lattices and the finite energy in-gap modes in topological insulators. We envision the presented mathematical and numerical framework to be invaluable for many technological advances pertaining to wave phenomenon such as reconfigurable waveguide designs.
\end{abstract}
\maketitle


\noindent{\textit{Introduction.}}---In the past few years, the concept of topological mechanical/elastic systems has led to a variety of intriguing development \cite{KaneLubensky2013,lubensky2015phonons,Wang2015,Khanikaev2015TopologicallyLattice,he2016acoustic_bandinv2,NiXiang2017,Rocklin2017,mao2018maxwell,MA2018FiniteFreq,chen2018topological,Chen2018ElasticLattices,ma2019topological,Ma2019QVHE,liu2019experimental,Xue2019AcousticLattice,SunMao2020,Wu2020,Danawe2021PRB}. In analogy to topological states in quantum many-body systems, the nontrivial topology structure from phonon bands grants these materials novel properties such as topologically protected edge/surface/corner modes. In general, current studies about topological mechanical/elastic systems can be classified into two categories. In the first category, the dynamic matrix of a elastic system is mapped to the Hamiltonian of an electronic system. Utilizing topological classifications developed for electronic systems~\cite{Hasan2010Colloquium,Qi2011RMP,Kitaev2009,Schnyder2008Classification,Po2017indicators,Bradlyn2017TQC}, this mapping enables mechanical systems to achieve the same type of topological phenomena, such as topological edge states in quantum Hall (or spin-Hall or valley-Hall) insulators \cite{Wang2015,Khanikaev2015TopologicallyLattice,NiXiang2017,chen2018topological,Chen2018ElasticLattices,he2016acoustic_bandinv2,Xue2019AcousticLattice,Ma2019QVHE,ma2019topological,liu2019experimental,Wu2020}. The second category is known as Maxwell systems \cite{KaneLubensky2013,lubensky2015phonons,mao2018maxwell,MA2018FiniteFreq,SunMao2020}. For these systems, the nontrivial topology is not coded in the dynamic matrix. Instead, it focuses on the connection between elastic constraints and the degrees of freedom, which maps elastic problem into a superconductor, known as the BDI class~\cite{Kitaev2009,Schnyder2008Classification,KaneLubensky2013}. From there, topological indices can be defined, which governs zero-energy topological states at edges.

These two classes of topological mechanical systems involve totally different concepts and theoretical description. More importantly, they exhibit distinct topological phenomena. For topological systems in the first category,  the topological phenomenon have to manifest themselves as high frequency physics, i.e., the topological edge/surface/corner states can only arise between two phonon bands (above the acoustic bands), and fundamental physics principles prevent such topological states to emerge below the acoustic band. This is because acoustic band is the lowest phonon band, and thus if mapped to electrons, topological indices are required to be zero below the lowest available energy bands. For the second category, on the contrary, topological states must be at  (or close to) zero energy, which is below the lowest phonon bands, and fundamental physics principle prohibit such topological states to arise above the acoustic band. In other words,  these two classes of topological phenomena are separated in frequency by fundamental principles. There is also important difference between these two categories regarding the dispersion of edge modes. In the first category, topological edge modes are typically disperse (usually connect the bulk bands above and below the gap). In contrast, topological edge modes in Maxwell systems are dispersionless (i.e., they form flat bands).

Very recently, there arises a new progress in elasticity called mechanical duality where the mechanics of two apparently different physical systems is related via mathematical mappings. If the system maps onto itself, then is called self-dual and it shows remarkable properties. Recently, Fruchart et al. \cite{fruchart2020dualities} found that the elastic twisted kagome lattice show duality while transitioning through its collapse mechanism \cite{GUEST2003} where two different structural configurations, equidistant from a mechanical critical point, have same dynamic characteristics and related elastic moduli. At the critical point, the twisted kagome lattice is self-dual and has a two-fold degenerate dispersion band structure. Later, Gonella \cite{Gonella2020} numerically demonstrated the duality in twisted kagome lattices by stitching together two dual configurations forming a heterogeneous bi-domain structure. More recently, Danawe et al. \cite{Danawe2021PRB} observed peculiar (d-2)-dimensional in-gap corner modes in self-dual kagome lattice occurring at a finite in-gap frequency. 

In this Letter, we show that with the help of mechanical duality, a new type of topological mechanical system arises, which exhibit properties of both categories discussed above. Same as the first category, these topological states arises at high frequency above acoustic bands, in band gaps between various phonon bands. However, the origin and topological structure of these topological states follow the same principle as Maxwell systems, and the topological edge (or domain-wall) states are dispersionless.

\begin{figure}
\includegraphics{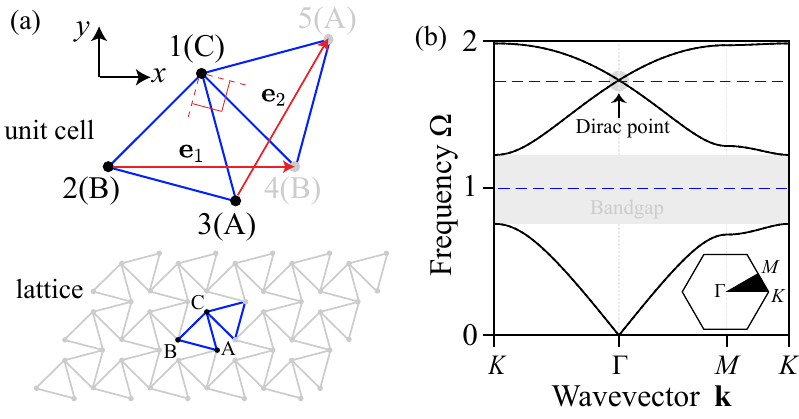}
\caption{\label{fig:unitcell} (a) A self-dual twisted kagome lattice and its unit cell with three equal masses at lattice sites A, B and C interconnected by bonds of stiffness $k$, $\textbf e_{1} $ and $\textbf e_{2} $ are the direct lattice basis vectors.  (b) The dispersion band structure of self-dual kagome lattice with all free lattice sites (solid lines) and pinned C lattice sites (dotted lines). The flat bands for lattice with pinned C sites appear at $\Omega=1$ (in the band gap of free lattice) and $\Omega=\sqrt3$ (at Dirac point of free lattice), where $\Omega=\omega{\sqrt{\frac{m}{k}}}$. The first irreducible Brillouin zone $K-\Gamma-M-K$ is shown in the inset.}
\end{figure}

\noindent{\textit{Self-dual kagome lattice.}}---A kagome lattice is characterized by three equal masses $m$ located at lattice cites A, B and C on the vertices of an equilateral triangle as shown in Fig. \ref{fig:unitcell}(a). The masses are interconnected by elastic bonds of stiffness $k$. The self-dual kagome lattice has same types of bonds of the unit cell oriented perpendicular to each other. For example, in Fig. \ref{fig:unitcell}(a), the two CA bonds are at 90$^{\circ}$ to each other, and similarly, the two CB bonds and two AB bonds are perpendicular to each other. The mass at each node can translate in the $x-$ and $y-$ directions and the displacement of the $\ell^{\text{th}}$ node can be represented by a 2D vector $\mathbf{u}^{{T}}_\ell = (u^x_\ell,  u^y_\ell)$, i.e., two degrees of freedom per node. By virtue of the periodicity, the displacements of nodes 2-4 and 3-5 are related and governed by Bloch's theorem, such that:
\begin{subequations}
    \begin{equation}
    \mathbf{u}_4 = \text{e}^{\textbf{i} \mathbf{k} \cdot \mathbf{e}_1}\mathbf{u}_2=\text{e}^{\textbf{i}q_1}\mathbf{u}_2
\end{equation}
\begin{equation}
    \mathbf{u}_5 = \text{e}^{\textbf{i} \mathbf{k} \cdot \mathbf{e}_2}\mathbf{u}_3=\text{e}^{\textbf{i}q_2}\mathbf{u}_3
\end{equation}
\label{eq:Bloch_conds}
\end{subequations}
where $\mathbf k$ is the Bloch wave vector, $\mathbf e_{1}$, $\mathbf e_{2}$ are direct lattice basis vectors and $q_{1}$, $q_{2}$ are reduced wave vectors given by $q_{1}=\mathbf k \cdot \mathbf e_{1}$, $q_{2}=\mathbf k \cdot \mathbf e_{2}$. Thus there are total six degrees of freedom (DOFs) per unit cell corresponding to the three nodes 1, 2 and 3. The dispersion band structure of a self-dual kagome lattice is shown in Fig. \ref{fig:unitcell}(b) having three doubly degenerate dispersion branches (solid lines) i.e., for every wave vector $\textbf k$ there are three pairs of identical eigenfrequencies. Now, if the C sites of the lattice are pinned, the unit cell is left with only 4 DOFs and the band structure reduces to two doubly degenerate flat bands as shown by dotted lines in Fig. \ref{fig:unitcell}(b) (see Supplemental Material for more details~\cite{SM2021}). Interestingly, the flat bands at $\Omega=1$ are in the band gap of the lattice with all free sites and that at $\Omega=\sqrt3$ pass trough the Dirac point of the free lattice band structure. For more details on band structure calculation of twisted kagome lattice as function of twist angle, see Ref. \cite{Danawe2021PRB}, where the author demonstrated existence of corner modes in a self-dual kagome lattice  which also evidently happen to appear at $\Omega=1$ characterized by zero deformation of same type of lattice sites, as if they are pinned. In this Letter, we further investigate the localized states near intentionally pinned sites of same type (A, B, or C) in the bulk of self-dual kagome lattice with the reason for their existence and their topological nature.

\noindent{\textit{Finite-frequency localized modes.}}---What will happen if some (but not all) of the $C$ sites are pinned? For such a partially pinned self-dual kagome lattice, it turns out that an intriguing phenomena emerge: no matter how many C sites we choose and regardless of which $C$ sites are selected, each pinned C site always generates 4 modes localized around this site, two at frequency $\Omega=1$ and two at $\Omega=\sqrt{3}$ (see Supplemental Material for more details \cite{SM2021}). In a lattice system, localized modes induced by a pinned site is not uncommon. However, if we pin two (or more) sites close to each other, these localized modes will typically hybridize with each other and thus their frequency shall shift depending on the distance between these pinned sites. Such hybridization never arise in the self-dual kagome lattice, and the frequency of these localize mode always remains exact $\Omega=1$ or $\sqrt{3}$, even if two pinned C sites are right next to each other. This absence of hybridization is a unique property of this self-dual lattice, and is one of the key results of this study. 

In addition, these localized modes also have some other intriguing properties. Firstly, although only some of the C sites are pinned, for all these $\Omega=1$ or $\sqrt{3}$ modes, all C sites in the entire lattice exhibit zero displacement (i.e., all C sites are effectively pinned) similar to the corner modes observed in Ref. \cite{Danawe2021PRB}. Secondly, this phenomenon is extremely robust and doesn't exhibit any finite-size or boundary effect. The same phenomena and exact frequencies are observed regardless of system size (from a few unit cells to infinite lattices) or boundary conditions (open or periodic). The location of the pinned sites (near the edge or in the bulk) has no impact either.

\begin{figure}
     \centering
\includegraphics[]{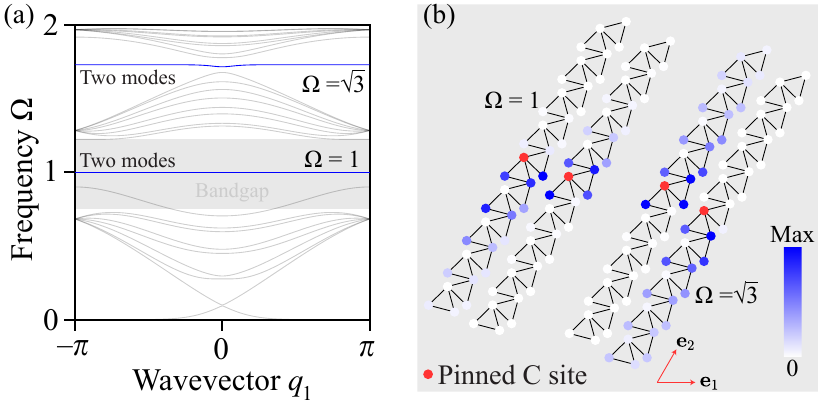}
     \caption{(a) Eigenfrequencies of a supercell with a pinned lattice site in the bulk. The two doubly degenerate flat bands appear at  $\Omega=1$ and $\Omega=\sqrt3$. (b) The mode shapes corresponding to the flat bands at $\Omega=1$ and $\Omega=\sqrt3$ localized near pinned lattice site for $q_1=2\pi/10$. The two modes, at the same frequency, decay away from the pinned lattice site in opposite directions with the same decay rate.}
     \label{fig:supercell}
\end{figure}

Because these localized modes never hybridize with each other, we can use them as building block to create more complicated structures. For example, if we pin one row of C sites along a straight or zigzag line, these localized modes will form a 1D waveguide, with four 1D flat bands, two at $\Omega=1$ and two at $\Omega=\sqrt{3}$. If two rows of C sites are pinned, two such waveguides are obtained. Even if the two waveguides are very close to each other, the waveguide modes would not hybridize between the two waveguides. If we pin all the C sites, these localized modes produces four 2D flat bands as shown in Fig.~\ref{fig:unitcell}(b). To better demonstrate this effect, in Fig.~ \ref{fig:supercell}(a), we present the phonon band structure with one row of C sites pinned down, calculated using the supercell shown in Fig.~\ref{fig:supercell}(b). Two flat 1D bands at $\Omega=1$ and two at $\Omega=\sqrt{3}$ are obtained. These modes are localized near the row of pinned C sites (except at $q_{1}=0$, $\Omega=\sqrt{3}$) with exponentially decaying mode shape away from the pinned sites.

\noindent{\textit{\label{analytical_formulation}Topology and analytic theory.}}---
It turns out that these robust features have the same topological origin as the zero-frequency topological edge modes in Maxwell systems, i.e., a topological winding number from the Maxwell counting argument~\cite{KaneLubensky2013,lubensky2015phonons,mao2018maxwell}. However, because the topological modes here are at finite frequencies, a new tool of localized basis needs to be introduced.

In a lattice system, any deformation can be characterized by the displacement field $\mathbf{W} =(\mathbf{u}_1^T, \mathbf{u}_2^T, \ldots ,\mathbf{u}_{N_s}^T)^T$, where $\mathbf{u}_i$ is the deformation vector of the $i$th lattice site. This deformation vector has $d \times N_s$-components, where $d$ is the space dimension and $N_s$ is the number of sites. We define two special sets of deformation fields, $\mathbf{W}^{+}_{\langle i,j\rangle}$ and $\mathbf{W}^{-}_{\langle i,j\rangle}$, which will serves as basis of our topological modes. 
Here, $\langle i,j\rangle$ represent a bond connecting two neighboring sites $i$ and $j$. For the deformation $\mathbf{W}^{+}_{\langle i,j\rangle}$, all other lattice sites exhibit zero displacement, except sites $i$ and $j$, which share the same displacement vector, $\mathbf{u}_i = \mathbf{u}_j = \mathbf{n}_{\langle i,j\rangle}$ with $\mathbf{n}_{i,j}$ being the unit vector along the bond $\langle i,j\rangle$. For $\mathbf{W}^{-}_{\langle i,j\rangle}$, it is very similar except that $i$ and $j$ have opposite displacements $\mathbf{u}_i = - \mathbf{u}_j = \mathbf{n}_{\langle i,j\rangle}$.

Here, we focus on symmetric deformations $\mathbf W^{+}$, which gives eigenmodes at $\Omega=1$. The anti-symmetric ones $\textbf W^{-}$ follow exactly the same physics, and they produce eigenmodes at $\Omega=\sqrt{3}$. Using symmetric deformations $\mathbf W^{+}$, we can construct the following displacement field
\begin{align}
    \mathbf{W}_{\textrm{AB}}= \sum_{\langle \textrm A_i,\textrm B_j\rangle} \mathcal{A}_{\langle \textrm A_i, \textrm B_j\rangle} \mathbf{W}^{+}_{\langle \textrm A_i,\textrm B_j\rangle}
    \label{eq:BC_bond_modes}
\end{align}
This deformation is a linear superposition of $\mathbf W^{+}$, and $\mathcal{A}_{\langle \textrm A_i,\textrm B_j\rangle}$ is the coefficient/amplitude for each $\mathbf W^+$. Here, we only use bonds connecting a A site and a B site, and therefore all C sites have zero deformation.
Similarly, we can define $\mathbf{W}_{\textrm {CA}}$ or $\mathbf{W}_{\textrm{BC}}$ using CA or BC bonds, respectively. Here, we shall focus on $\mathbf{W}_{\textrm{AB}}$, and the same results can be easily generalized to $\mathbf{W}_{\textrm{CA}}$ and $\mathbf{W}_{\textrm{CB}}$.

\begin{figure}
     \centering
\includegraphics{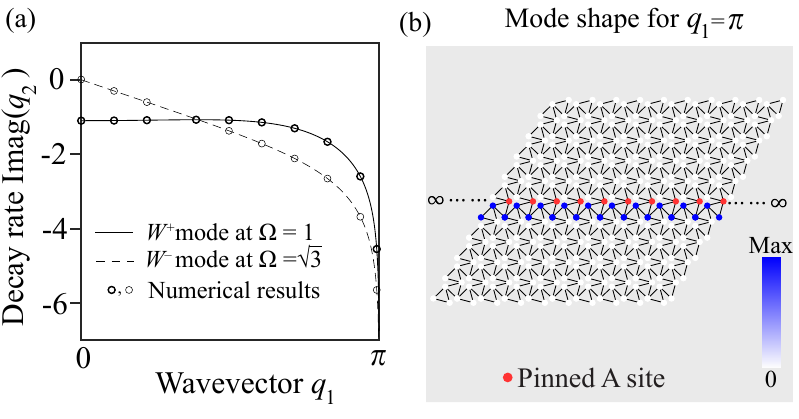}
     \caption{(a) The decay rate of edge modes obtained from the compatibility matrix formulation compared with the decay rate from supercell simulations. (b) The mode shape of a infinite ribbon with pinned row of A lattice sites for $q_1=\pi$ at which the decay rates approach $-\infty$ resulting in the highly localized edge mode near the the pinned row of lattice sites.}
     \label{fig:decay}
\end{figure}

In general, $\mathbf{W}_{\textrm{AB}}$ is not an eigenmode of the dynamic matrix. However, it is straightforward to verify that for the self-dual lattice, $\mathbf{W}_{\textrm{AB}}$ becomes an eigenmode with frequency $\Omega=1$, if the following constraint is obeyed: all C sites stay at their equilibrium positions (pinned or at force balance). Therefore, to study the $\Omega=1$ modes, we can use the linear space of $ \mathbf{W}_{\textrm{AB}}$, where the number of degrees of freedom is the number of AB bonds $N_{dof}=N_{\textrm{AB}}$. At the same time, without pinning, the total constraint number is $N_{c}=2 N_{\textrm C}$, because the $x$ and $y$ components of the total force on each C site need to remain zero. Remarkably, for a kagome lattice, these two numbers coincide, $N_{dof}=N_c$, and thus the system is at the Maxwell point.

Same as in topological mechanics, here we can define an effective compatibility matrix to connect the degrees of freedom and the constraints.
\begin{align}
\mathbf{F}=\mathbf{C}_{\textbf{eff}} \mathbf{\mathcal{A}}
\label{eq:effective:cmatrix}
\end{align}
Here, $\mathbf{F}=(F_{1,x}, F_{1,y}, F_{2,x},F_{2,x} \ldots)^T$ is a $N_c$ component vector, where $F_{i,x}$ and $F_{i,y}$ are the $x$ and $y$ components of the total force on the $i$th C site. $\mathbf{\mathcal{A}}$ is an $N_{dof}$ dimensional vector composed of the coefficients $\mathcal{A}$ in Eq.~\eqref{eq:BC_bond_modes}. 

In analogy to Maxwell topological mechanics, the null-space of the $\mathbf{C}_{\textbf{eff}}$ matrix (i.e. all modes obeying $\mathbf{C}_{\textbf{eff}}\mathbf{\mathcal{A}}=0$) corresponds to $\textbf W^+$ modes at $\Omega=1$. For a lattice with periodic boundary condition and without any pinning sites, $N_c=N_{dof}$, and thus $\mathbf{C}_{\textbf{eff}}$ is a square matrix. As shown in the Supplemental Material~\cite{SM2021}, here $\det \mathbf{C}_{\textbf{eff}} \ne 0$, and thus the null-space is empty, indicating the absence of any $\Omega=1$ modes. However, once some C sites are pinned, $\mathbf{C}_{\textbf{eff}}$ is no longer a square matrix. Instead, the number of degrees of freedom now exceeds the number of constrains $N_{dof}>N_c$, and thus the null space shall contain $N_{dof}-N_c$ independent modes. It is easy to realize that for every pinned C site, $N_c$ reduces by $2$ and thus $N_{dof}-N_c$ increases by $2$. This the reason why we obtains two $\Omega=1$ modes for every pinned C sites. The same approach and conclusions also apply to $\textbf W^-$ modes at $\Omega=\sqrt3$, except that we have bulk $\textbf W^-$ modes at zero wave-vector corresponding to the Dirac point.

Same as in Maxwell topological mechanics, a topological index can be defined for this $\mathbf{C}_{\textbf{eff}}$ matrix, which dictates the number of topologically protected edge/domain-wall modes~\cite{KaneLubensky2013,lubensky2015phonons,mao2018maxwell,SunMao2020}. To define this index, we need to switch to the momentum space, where the $\mathbf{C}_{\textbf{eff}}$ becomes (See Supplemental Material~\cite{SM2021})
\begin{align}
    \mathbf{C}_{\textbf{eff}}
   =  k\begin{pmatrix}
        \frac{1}{2} +\frac{3}{4}(e^{\textbf{i} q_1}+e^{\textbf{i} q_2})&  \frac{\sqrt{3}}{4}(e^{-\textbf{i} q_1}-e^{-\textbf{i} q_2})\\
            -\frac{\sqrt{3}}{4}(e^{\textbf{i} q_1}-e^{\textbf{i} q_2})& \frac{1}{2}+\frac{3}{4}(e^{-\textbf{i} q_1}+e^{-\textbf{i} q_2})
    \end{pmatrix}
\end{align}
For each value of $q_1$, a topological winding number can be defined as
\begin{align}
    n=\oint \frac{d z}{2\pi i} \textrm{tr} \left(C^{-1} \partial_z C\right)
\end{align}
where $z=e^{i q_2}$. Using the gauge-invariant integral contour introduced in Ref.~\cite{SunMao2020}, (i.e., the unit circle on the complex $z$ plane and remove the residue at $z=0$ or $z=\infty$), we can obtain two integer topological indices. For a line of pinned C sites (Fig.~\ref{fig:supercell}), at each $q_1$, these two topological indices dictates the number topologically-protected modes localized above and below the pinned line respectively (i.e., with a negative or positive decay rate). For the $\mathbf{C}_{\textbf{eff}}$ matrix here, both the two indices are unity, which means that for each $q_1$, we have two modes at $\Omega=1$ localized near this 1D line, one above and one below, in full agreement with numerical simulations.

In addition to the number of modes, the $\mathbf{C}_{\textbf{eff}}$ matrix also dictates their localization length and mode shape, same as Maxwell zero mode~\cite{Sun2012,KaneLubensky2013,lubensky2015phonons,Rocklin2017,mao2018maxwell,SunMao2020}. For a given $q_1$, the equation $\det \mathbf{C}_{\textbf{eff}}=0$ has a complex $q_2$ solution, and its imaginary part is the decay rate
\begin{align}
    \textrm{Im} q_2=\ln  \left (\frac{14+6\cos q_1 -\sqrt{142+96 \cos q_1+18 \cos 2 q_1}}{12 \cos\frac{q_1}{2}} \right) \label{decayrate1}
\end{align}
As shown in Fig.~\ref{fig:decay},
this analytic prediction perfectly agrees with the decay rated measured from supercell simulations.

\begin{figure}
     \centering
\includegraphics{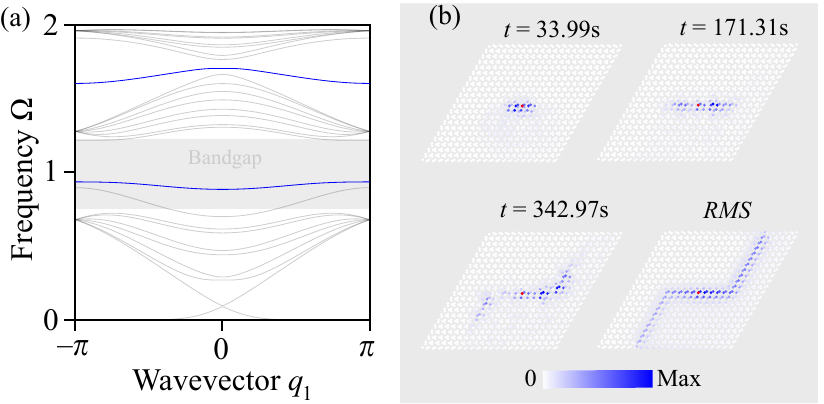}
     \caption{(a) Eigenfrequencies of a supercell with a loosely pinned lattice site. The flat bands of supercell with pinned lattice site become disperse due to loose pinning, but they remain two-fold degenerate.  (b) Wave propagation along the loosely pinned sites in a finite lattice due to non-zero group velocity. The central unit cell with a pinned lattice site is excited using a harmonic excitation and the displacement field is obtained as a function of time.}
     \label{fig:waveguide}
\end{figure}

\noindent{\textit{Loosely pinned waveguides.}}--- Instead of complete pinning, loosely pinning the lattice sites using an elastic foundation of finite spring stiffness  (here $4k$) results in eigenfrequency solutions of supercell as depicted in Fig. \ref{fig:waveguide}(a). The flat bands appearing in the band gap of supercell with pinned lattice sites (Fig. \ref{fig:supercell}(a)) are not flat in case of supercell with loosely pinned sites; however, they are still two-fold degenerate. With the loosely pinned sites, the non-zero group velocity allows transmission of wave energy along the pinned lattice sites whereas the bulk of the lattice remains isolated due to the band gap. The time snapshots and RMS of displacement field in a finite self-dual kagome lattice with loosely pinned lattice sites is shown in Fig. \ref{fig:waveguide}(b) proving the selective wave propagation along a desired path. The loosely pinned waveguide is reconfigurable by simply pinning and unpinning of lattice sites. 

The spatial decay of the two degenerate edge modes is on the opposite sides of the pinned row of lattice sites. Thus, exciting only one of the modes results in decay of edge modes only on one side of the waveguide that would completely isolate the other half of the finite lattice divide by the waveguide. For instance, the zigzag waveguide shown in Fig. \ref{fig:waveguide}(b) is excited at a point shown in red lying in the lower half of finite lattice. Thus, the edge modes decay away from the waveguide into the lower half. Hence, having a neighboring row of loosely pinned lattice in the upper half of finite lattice would result in zero interference between the two waveguides. 

\noindent{\textit{Conclusions.}}---In this work, we analyzed a new type of topological states in a self-dual kagome lattice which exist at two specific frequencies $\Omega=1,\sqrt3$ localized near pinned sites of a sublattice. These states appear at Maxwell point where the number of degrees of freedom is equal to number of constraints. Although analogous to topological mechanics in Maxwell lattices,  the Maxwell relation obtained for self-dual kagome lattice is fundamentally different and the modes are at finite frequency instead of zero frequency floppy modes, but they retain their dispersionless (flat band) behavior. These modes exhibit special deformation fields which are characterized by equal deformation of two lattice sites along the bond connecting them while the deformation of rest of the sites is zero. For a row of pinned sites of a sublattice, the topological modes are localized near the pinned sites while decaying exponentially in the bulk. The decay rate is obtained from the determinant of effective compatibility matrix and it is compared with supercell simulations with excellent agreement. The topological index for these modes is same as that for zero frequency modes in Maxwell lattices and it corroborates the existence of two topological modes at frequencies $\Omega=1$ and $\sqrt 3$.

\begin{acknowledgements} 
This work was supported in part by the National Science Foundation, grant number CMMI-1914583 (H.D. S.T.) and the Office of Naval Research MURI N00014-20-1-2479 (H.L. K.S.).
\end{acknowledgements}


\bibliography{main_text}

\end{document}


\title{Finite-Frequency Topological Maxwell Modes in Mechanical Self-Dual Kagome Lattices\\
(Supplemental Material)}

\author{Hrishikesh Danawe}
\affiliation{Department of Mechanical Engineering, University of Michigan, Ann-Arbor, MI 48109-2125, United States}
\author{Heqiu Li}
\affiliation{Department of Physics, University of Michigan, Ann-Arbor, MI 48109-2125, United States}
\affiliation{Department of Physics, University of Toronto, Toronto, Ontario, Canada}
\author{Kai Sun}%
\affiliation{Department of Physics, University of Michigan, Ann-Arbor, MI 48109-2125, United States}
\author{Serife Tol}
   
\affiliation{Department of Mechanical Engineering, University of Michigan, Ann-Arbor, MI 48109-2125, United States}

\maketitle

\section{Localized states near pinned sites in a Finite Lattice}
\begin{figure}
     \centering
\includegraphics{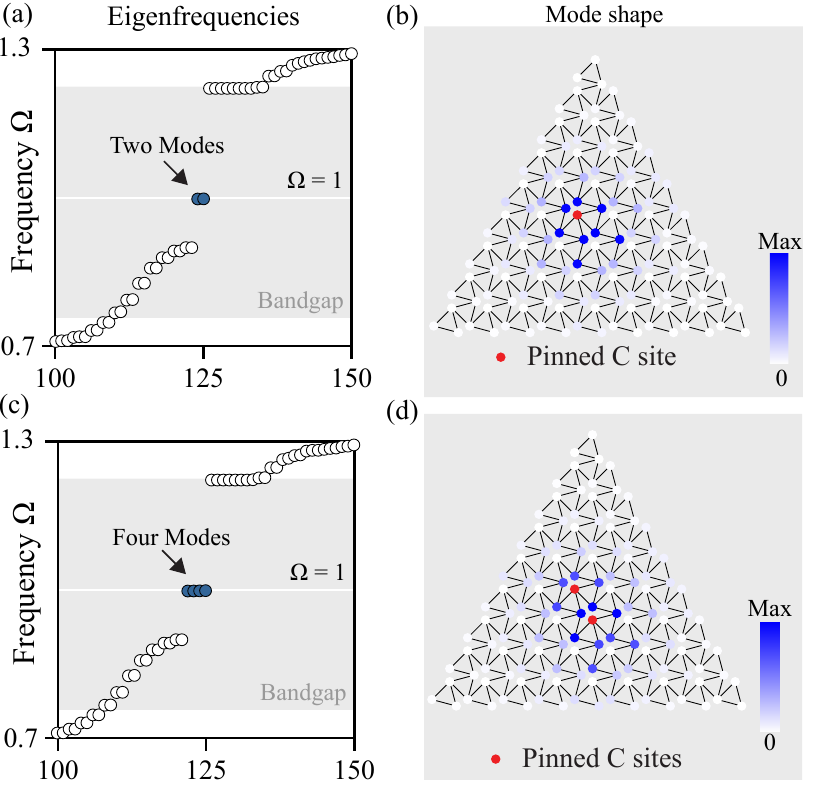}
     \caption{ (a) Eigenfrequencies of finite triangle-shaped self-dual kagome lattice with one pinned lattice site in the bulk. (b) Mode shape of localized states near the pinned site at frequency $\Omega=1$. (c) Eigenfrequencies of finite triangle-shaped self-dual kagome lattice with two pinned lattice site in the bulk. (d) Mode shape of localized states near the two pinned sites at frequency $\Omega=1$.}
     \label{fig:triangle}
\end{figure}

From the supercell analysis presented in the main text, it is evident that pinning a row of same lattice sites in bulk results in localized modes near the pinned sites. In this section, we pin one or multiple sites of the same kind in a finite lattice which has free boundary conditions for all truncated edges. Eigenfrequency solutions for a finite self-dual triangle-shaped lattice with one pinned C site is depicted in Fig. \ref{fig:triangle}(a) with two in-gap modes at $\Omega=1$. The mode shape corresponding to these modes is as shown in Fig. \ref{fig:triangle}(b). The mode is localized around the pinned C site and all C sites in the bulk have zero deformation. Apart from these modes, there are two more modes at $\Omega=\sqrt3$ localized near the pinned site in the bulk.  Now, if two of the C sites are pinned in the bulk, there exist four in-gap modes at $\Omega=1$ as shown in Fig. \ref{fig:triangle}(c). The mode shape of these modes is depicted in Fig. \ref{fig:triangle}(d). Similarly, there exist four modes at the frequency $\Omega=\sqrt3$ localized near the pinned sites. In conclusion, there are always two localized states per pinned site in the bulk at frequency $\Omega=1$ and $\sqrt3$ if all pinned sites are of same type (A, B or C) no matter how far or close these sites are located in the bulk. Also, if two sites of different type are pinned in the bulk, as long as they do not affect each other (i.e. the localized modes at one site decay before it reaches to other type of pinned site), there still exist two localized modes per pinned site at $\Omega=1$ and $\sqrt3$.  

\begin{figure}
     \centering
\includegraphics{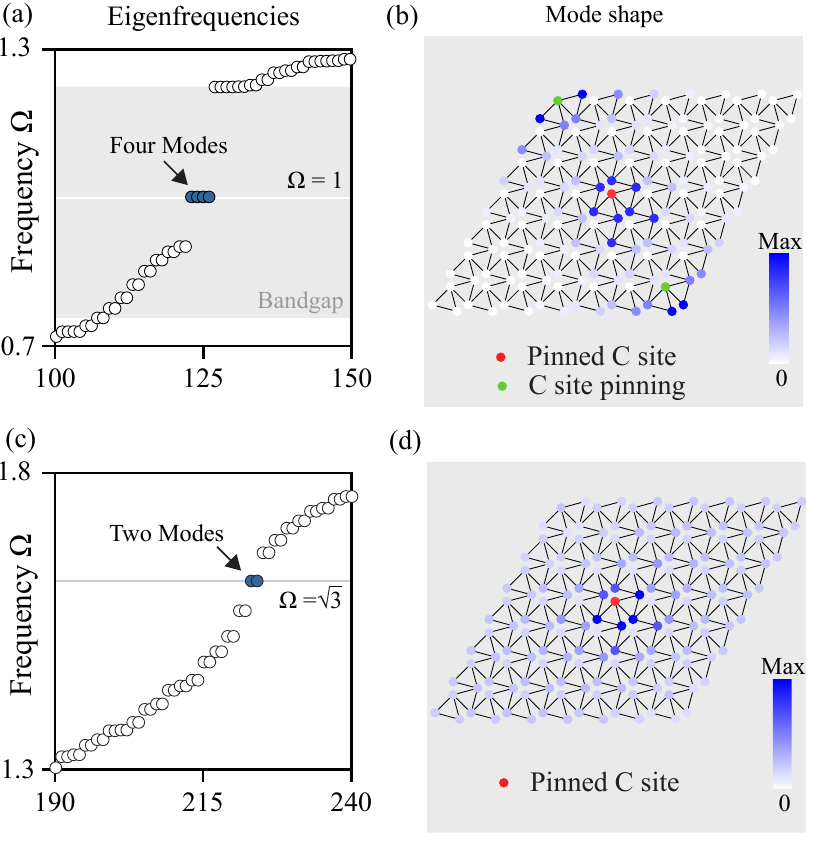}
     \caption{(a) Eigenfrequencies of finite parallelogram-shaped self-dual kagome lattice with one pinned lattice site in the bulk. (b) Mode shape of localized states near the pinned site and at the 120 deg angled corner at $\Omega=1$.(c) Eigenfrequencies of finite parallelogram-shaped self-dual kagome lattice with one pinned lattice site in the bulk. (d) Mode shape of localized states near the pinned site at $\Omega=\sqrt3$.}
     \label{fig:parallelogram}
\end{figure}
For a parallelogram-shaped self-dual kagome lattice with one pinned site in the bulk, there are four in-gap modes at $\Omega=1$ as shown in Fig. \ref{fig:parallelogram}(a) contrary to only two modes per site in a triangle-shaped lattice. The additional two modes corresponds to the corner modes localized at the two 120$^{\circ}$ angled corners that hold C sites which act as if they are pinned \cite{Danawe2021PRB}. Thus, the mode shape of these modes show localization near the single pinned site in the bulk and additionally at the 120$^{\circ}$ angled corners of the parallelogram.  Apart from these modes, there are two modes at $\Omega=\sqrt3$ for the pinned site in the bulk as shown in Fig.\ref{fig:parallelogram}(c) but no localization is observed near corners of the finite lattice at this frequency. Thus, the mode shape of modes at frequency $\Omega=\sqrt3$ are localized at the single pinned site in the bulk as depicted in Fig.\ref{fig:parallelogram}(d). 

\section{Self-Dual kagome Lattice with Pinned sublattice}
\begin{figure}
     \centering
\includegraphics{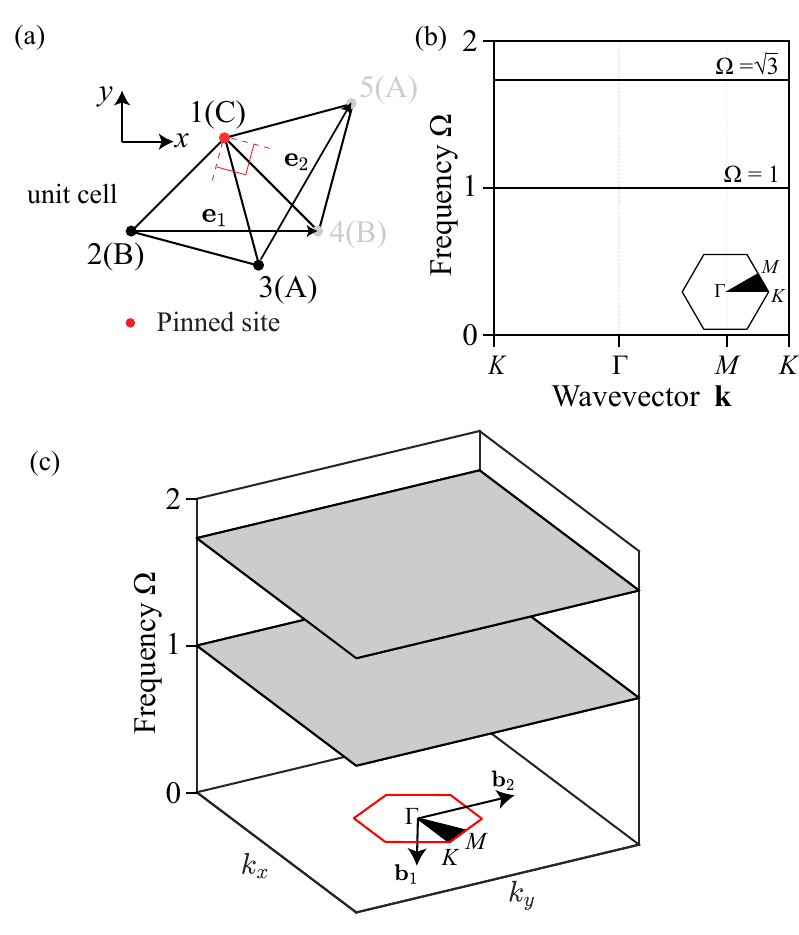}
     \caption{(a) Unit cell of self-dual twisted kagome lattice with a pinned sublattice. (b) The dispersion band structure of self-dual twisted kagome lattice with a pinned sublattice along the boundary of first irreducible Brillouin zone. (c) Dispersion surfaces of the same lattice showing two doubly degenerate flat bands with the first irreducible Brillouin zone depicted in inset.}
     \label{fig:pinned_sublattice}
\end{figure}

Figure \ref{fig:pinned_sublattice}(a) depicts the unit cell of a self-dual twisted kagome lattice with one pinned site, e.g. site C. The unit cell now constitutes only four degrees of freedom corresponding to the sites A and B. The dispersion band structure has two degenerate flat bands with zero group velocity as shown in Fig. \ref{fig:pinned_sublattice}(b). Thus, the frequencies of these modes take very specific values: $\Omega = 1$ and $\sqrt{3}$ for the lower two and upper two modes, respectively. The dispersion surfaces are depicted in Fig. \ref{fig:pinned_sublattice}(c) which are flat bands ($\Omega$ is independent of wavevector $\mathbf k$) at $\Omega = 1$ and $\sqrt{3}$. Pinning sublattice A or B, instead of sublattice C, results in identical dispersion bands.

The flat lines appearing in the bulk band gap of the self-dual kagome lattice are related to the flat dispersion bands of a lattice with one pinned site. Since the observed localized modes when a site C in the bulk is pinned, for instance, all bulk C sites have zero deformations for these localized modes while no pinning is enforced in the bulk (i.e., they are effectively “pinned”). In this sense, these modes are “descendants" of the flat bands obtained from the unit cell analysis of the pinned lattice. Interestingly, these modes can be solved analytically for the finite structure (very much similar to the analytical solution of the pinned flat bands), which can be shown to have eigenvalues at exactly $\Omega = 1, \sqrt3$.

\section{\label{analytical_formulation}spatially compact localized states}
The number of degrees of freedom is the number of AB bonds $N_{dof}=N_{\textrm{AB}}$. This counting comes from Eq. (2) in the main text, which contains $N_{\textrm{AB}}$ independent free parameters ($\mathcal{A}_{\langle \textrm A_i,\textrm B_j\rangle}$). In a kagome lattice, we have two AB bonds per unit cell, and thus $N_{dof}=N_{\textrm{AB}}=2 N$, where $N$ is the number of unit cells.
As for the number of constraint, for each C site, we have two constraints: the x- and y- components of the total force is zero. Thus, for the whole lattice, the number of constraint is $N_{c}=2 N_{\textrm C}$, where $N_{\textrm C}$ is the number of C sites. For a kagome lattice, we have one C site per unit cell, and thus $N_c=2 N_{\textrm C}=2 N$, where $N$ is the number of unit cells. Remarkably, here, we have exactly the same number of constrains and degrees of freedom $N_{dof}=N_{c}$, i.e., the system is at the Maxwell point.
It must be emphasized, in contrast to topological mechanics in Maxwell lattices, this Maxwell relation we found here is fundamentally different. Most importantly, as will be shown below, the counting argument here deals with finite frequency modes instead of zero energy ones.

Here we consider deformations defined in Eq. (2) in the main text and the deformation of C sites are set to zero.
Same as in topological mechanics, we can define an effective compatibility matrix to connect the degrees of freedom and the constraints given by Eq. (3) in the main text.

For a lattice with periodic boundary condition and without any pinning sites, $N_c=N_{dof}$, and thus $\mathbf{C}_{\textbf{eff}}$ is a square matrix. As will be shown below for $\textbf W^+$ modes (at frequency $\Omega=1$), this matrix has a nonzero determinant. Therefore, to satisfy the constraint $\mathbf{F}=0$, there is only a trivial solution $\mathbf{\mathcal{A}}$, indicating the absence of $\Omega=1$, $\textbf W^+$ modes. However, once some C site is pinned, $\mathbf{C}_{\textbf{eff}}$ will no longer be a square matrix. Instead, the number of degrees of freedom ($N_{dof}$) now exceeds the number of constrains $N_c$, and thus we shall find $N_{dof}-N_c$ independent nontrivial solutions for the constraint $\mathbf{F}=0$, which gives us $N_{dof}-N_c$ independent modes at frequency $\Omega=1$. This number $N_{dof}-N_c$ is precisely twice the number of pinned C sites. In other words, for each pinned C sites, we shall obtain two modes at $\Omega=1$.
The same approach and conclusions also follows to $\textbf W^-$ modes at $\Omega=\sqrt3$, except that we have bulk $\textbf W^-$ modes at zero wave-vector corresponding to the Dirac point.

It is worthwhile to emphasize that these modes induced by site pinning has a finite localization length. In other words, these modes are not due to local motions of one or a few cites near the pinned site. Instead, its wavefunction extends away from the pinned sites, and the amplitude decays gradually as we move away from the pinned site. Typically, if two localized modes with the same frequency are brought together, hybridization between the two modes will lift the degeneracy, splitting one frequency into two. But for the topological modes here, even if we pin two neighboring C sites, where the $\Omega=1$ or $\sqrt{3}$ modes from the two sites have strong real-space over lap, their frequency will stay at the same value without any splitting. This is one unique feature and a consequence of topological protection.

To better understand these topological modes and to define the topological index, here, we transfer from the real space to the $k$-space, utilizing Bloch's theorem, which requires the deformation to obey Bloch conditions (Eq. (1) in the main text).
We use site labels shown in Fig. 1(a) in the main text, where site $1$ belongs to the C sublattice and sites $2$ and $4$ ($3$ and $5$) belong to the B(A) sublattice. For a given wavevector, Bloch modes composed of $\textbf W^+$ modes can be written as
\begin{align}
    \mathbf{u}_{2}(q_1,q_2)=\mathcal{A}_1  \mathbf{n}_{\langle 2,3 \rangle} + \mathcal{A}_2  \mathbf{n}_{\langle 4,5 \rangle} \text{e}^{-\textbf{i}q_1}
    \\
    \mathbf{u}_{3}(q_1,q_2)=\mathcal{A}_1  \mathbf{n}_{\langle 2,3 \rangle} + \mathcal{A}_2  \mathbf{n}_{\langle 4,5 \rangle} \text{e}^{-\textbf{i}q_2}
    \label{eq:BC_bond_Bloch_modes}
\end{align}
Because we have two AB bonds per unit cell (between sites 2 and 3, and between sites 4 and 5), each bonds may contribute a different amplitude, which is represented by $\mathcal{A}_1$ and $\mathcal{A}_2$ respectively in this formulae.
$\mathbf{n}_{\langle 2,3 \rangle}$ and $\mathbf{n}_{\langle 4,5 \rangle}$ are unit vectors along the bond 2-3 and 4-5, respectively.
Deformation for site 4 and 5 can be obtained via the Bloch conditions (Eq. (1) in the main text), and thus we can obtain total force applied to site $1$ as
\begin{align}
\mathbf{F}= \sum_{i=2}^{5}k(\mathbf{n}_{\langle i,1 \rangle} \cdot \mathbf{u}_i)\; \mathbf{n}_{\langle i,1 \rangle}
\label{eq:force:site:1}
\end{align}
where $\mathbf{n}_{\langle i,1 \rangle}$ is the unit vector along bond $\langle i,1 \rangle$ and $\mathbf{F}=(F_x,F_y)$ is the force on site 1.
This relation can be written in a matrix form
\begin{align}
    \begin{pmatrix}
    F_x\\ F_y
    \end{pmatrix}
=
    \mathbf{C}_{\textbf{eff}}
     \begin{pmatrix}
   \mathcal{A}_1\\ \mathcal{A}_2
    \end{pmatrix}
\end{align}
where $\mathbf{C}_{\textbf{eff}}$ is a $2\times 2$ matrix. It is the $k$-space version of the same effective compatibility matrix defined above. For the self-dual kagome lattice
\begin{align}
    \mathbf{C}_{\textbf{eff}}
   =  k\begin{pmatrix}
        \frac{1}{2} +\frac{3}{4}(e^{\textbf{i} q_1}+e^{\textbf{i} q_2})&  \frac{\sqrt{3}}{4}(e^{-\textbf{i} q_1}-e^{-\textbf{i} q_2})\\
            -\frac{\sqrt{3}}{4}(e^{\textbf{i} q_1}-e^{\textbf{i} q_2})& \frac{1}{2}+\frac{3}{4}(e^{-\textbf{i} q_1}+e^{-\textbf{i} q_2})
    \end{pmatrix}
\end{align}

In this $k$-space formula, at each wave-vector, there are two degrees of freedom ($\mathcal{A}_1$ and $\mathcal{A}_2$) and two constraints ($F_x=0$ and $F_y=0$). Thus, we are the Maxwell point as discussed early on. The $\mathbf{C}_{\textbf{eff}}$ matrix here connects degrees of freedom with constraints, and thus it serves the same role as the compatibility matrix in topological mechanics. Therefore, we will name this matrix the effective compatibility matrix.

This effective compatibility matrix is a function of $e^{\textbf{i} q_1}$ and $e^{\textbf{i} q_2}$. In an infinite lattice (with periodic boundary conditions), the system is at the Maxwell point, and it is easy to verify that $\det \mathbf{C}_{\textbf{eff}} \ne 0$ for any real $q_1$ and $q_2$. Thus, in such a setup, the constraint $\mathbf{F}=0$ only has a trivial solution $\mathcal{A}_1=\mathcal{A}_2=0$, i.e., no $\Omega=1$ mode is expected. This is consistent with the phonon band structure, which shows a gap around $\Omega=1$.

If we pin one row of C sites along a straight line parallel to $\mathbf{e_1}$, the transnational symmetry along the $\mathbf{e_1}$ direction implies that $q_1$ is still a well-defined vector. Here, in order to obtain nontrivial solutions of the constraints $\mathbf{F}=0$, we must require $\det \mathbf{C}_{\textbf{eff}}=0$. For a given $q_1$, a complex solution of $q_2$ can be found
\begin{align}
    q_2=\frac{q_1}{2} -\textbf{i} \ln \frac{-14-6\cos q_1 \pm \sqrt{142+96 \cos q_1+18 \cos 2 q_1}}{12 \cos\frac{q1}{2}} 
\end{align}
These waves with a real $q_1$ and a complex $q_2$ are localized 1D modes near the pinned 1D array of sites. The imaginary part of $q_2$ is the decay rate, which is
\begin{align}
    \ln  \left (\frac{14+6\cos q_1 -\sqrt{142+96 \cos q_1+18 \cos 2 q_1}}{12 \cos\frac{q_1}{2}} \right) \label{decayrate1}
\end{align}
As we move away from the pinned 1D array, for each lattice constant $\mathbf{e_2}$, the amplitude of deformation decreases by a factor of 
\begin{align}
    \left |\frac{-14-6\cos q_1 \pm \sqrt{142+96 \cos q_1+18 \cos 2 q_1}}{12 \cos\frac{q_1}{2}} \right|
\end{align}

These localized modes are topologically protected. The topological index and bulk-edge correspondence are in strong analogy to the zero-energy topological modes defined in Maxwell lattices.
To define the topological index, we utilize the determinant of the $\mathbf{C}_{\textbf{eff}}$ matrix. For each given $q_1$, $\det \mathbf{C}_{\textbf{eff}}$ is a function of $e^{\textbf{i} q_2}$. We can define a complex coordinate  $z_2=e^{\textbf{i} q_2}$. As shown in Ref.\cite{SunMao2020}, $\det \mathbf{C}_{\textbf{eff}}$ is a polynomial function $z_2$ (negative powers of $z_2$ is allowed), and the number of localized mode on each side of the pinned 1D array of sites is dictated by the topological index  [Eq. 6 in Ref.\cite{SunMao2020} and the contour follows figure 1(c). T/B gives topological modes number above/below the 1D pinned line].

For $\textbf W^-$ modes, the compatibility matrix is
\begin{align}
    \mathbf{C}_{\textbf{eff}}
   =  k\begin{pmatrix}
       \frac{3}{4}(e^{\textbf{i} q_1}-e^{\textbf{i} q_2})&  \frac{\sqrt{3}}{2} -\frac{\sqrt{3}}{4}(e^{\textbf{i} q_1}+e^{\textbf{i} q_2})  \\
           - \frac{\sqrt{3}}{2} +\frac{\sqrt{3}}{4}(e^{-\textbf{i} q_1}+e^{-\textbf{i} q_2})  & \frac{3}{4}(e^{-\textbf{i} q_1}-e^{-\textbf{i} q_2}))
    \end{pmatrix}
\end{align}
and the decay rate (the imaginary part of $q_2$) is
\begin{align}
    \ln \left(\frac{3-\cos q_1-\sqrt{14-2 \cos q_1}\sin\frac{q_1}{2}}{2 \cos\frac{q_1}{2}}\right)\label{decayrate3}
\end{align}
 The decay rates given by Eq.\ref{decayrate1} and Eq.\ref{decayrate3} are plotted in Fig. 3(a) in the main text along with the decay rates obtained from supercell simulations for $q_1$ taking values in the range 0 to $\pi$. There is an excellent agreement between the theoretical and numerical results at both the frequencies $\Omega=1$ and $\sqrt3$. The decay rates at both the frequencies are found to approach $-\infty$ when reduced wave vector $q_1=\pi$. Thus, at this extreme condition, the mode shape is highly localized near the pinned site as depicted in Fig. 3(b) in the main text where only the unit cell sites closet to the pinned site show significant deformation whereas all other sites have negligible deformation.

\bibliography{sm.bib}

